\documentclass[showpacs,preprintnumbers,amsmath,amssymb]{revtex4}
\usepackage{graphicx}
\usepackage{dcolumn}
\usepackage{bm}

\begin{document}
\title{
Competing field pulse induced dynamic transition in Ising models
}

\author{Arnab Chatterjee}
\email{arnab@cmp.saha.ernet.in}
\author{Bikas K Chakrabarti}%
\email{bikas@cmp.saha.ernet.in}
\affiliation{
Saha Institute of Nuclear Physics\\
1/AF Bidhannagar, Kolkata 700 064, India.\\}

\begin{abstract}

The dynamic magnetization-reversal phenomena in the Ising model under a 
finite-duration external magnetic field competing with the existing order 
for $T<T_c^0$ has been discussed. The nature of the phase boundary has been
estimated from the mean-field equation of motion. The susceptibility and 
relaxation time diverge at the MF phase boundary. A Monte Carlo study
also shows divergence of relaxation time around the phase boundary.
Fluctuation of order parameter also diverge near the phase boundary.
The behavior of the fourth order cumulant shows two distinct behavior:
for low temperature and pulse duration region of the phase boundary the value 
of the cumulant at the crossing point for different system sizes is much less 
than that corersponding to the static transition in the same dimension which
indicate a new universality class for the dynamic transition. Also, for higher
temperature and pulse duration, the transition seem to fall in a mean-field
like weak-singularity universality class.

\end{abstract}
\pacs{05.50.+q; 05.70.Fh}
\maketitle

\section {Introduction}

\noindent 
Thin films of ferromagnetic materials are of considerable interest 
as potential materials for the use as ultra-high density recording 
medium. Recently, sophisticated experimental techniques such as the magnetic 
force microscopy have enabled physicists to look at the 
magnetization state and switching mechanism of ferromagnetic particles 
in the nanometre scale \cite{marwic}. Experiments \cite{bet,elm} on 
iron sesquilayers grown on tungsten are interesting from the theoretical 
point of view as the results seem to correspond precisely to an ideal 
ferromagnetic Ising system \cite{elm}. 

The response of a pure magnetic system to time-dependent external magnetic 
fields has been of current interest in statistical physics \cite{rmp,
korn1,pul2,acbkcpul}. These studies, having close applications in recording
and switching industry, have also got considerable scientific and technical 
interest. These spin systems, driven by time-dependent external magnetic fields,
have basically got a competition between two time scales: the time-period
of the driving field and the relaxation time of the driven system.
This gives rise to interesting non-equilibrium phenomena. T\'{o}me
and Oliveira first made a mean-field study \cite{tome} of kinetic
Ising systems under oscillating field. The existence of the dynamic
phase transition for such a system and its nature have been thoroughly
studied using extensive Monte-Carlo simulations \cite{korn1,mabkc1}. 
Recently, Junier and Kurchan \cite{jk} have shown that the dynamics under
oscillating field can induce asymmetry in the order parameter; namely 
an isotropic Heisenberg model (classical) under an alternating field switches
to XY-like or Ising-like behavior, as the external field frequency and
amplitude is tuned.

Later, investigations
were extended to the dynamic response of (ferromagnetic) pure Ising
systems under magnetic fields of finite-time duration \cite{pul1,pul1-1}.
All the studies with pulsed field were made below \( T^{0}_{c} \),
the static critical temperature (without any field), where the system
gets ordered. A `positive' pulse is one which is applied along the
direction of prevalent order, while the `negative' one is applied
opposite to that. The results for the positive pulse case did not
involve any new thermodynamic scale \cite{pul1}. In the negative pulse case, 
however, interesting features were observed 
\cite{acbkcpul,pul1-1,mish1,mish1-1,mc00,amth}:
the negative field pulse competes with the existing order, and the system
makes a transition from one ordered state characterised by an equilibrium
magnetisation \( +m_{0} \) (say) to the other equivalent ordered
state with equilibrium magnetisation \( -m_{0}, \) depending on the
temperature \( T \), field strength \( h_{p} \) and its duration
\( \Delta t \). This transition is well studied in the limit 
$\Delta t \rightarrow \infty$ for any non-zero value of $h_{p}$ at any 
$T<T_{c}^{0}$. This transition, for the general cases of finite \( \Delta t \),
is called here `magnetisation-reversal' transition. As in the oscillating
field case, in the study of magnetizatin-reversal transition, the finite 
duration ($\Delta t$) of the external field introduces a competition between 
two time scales: the relaxation time of the system and $\Delta t$.
Some aspects of this transition has recently been studied extensively 
\cite{pul2,acbkcpul,pul1-1,mish1,mish1-1}.
Recently, studies on switching behavior of some thin films of 
`perpendicular media' (where spins 
are Heisenberg like rather than Ising like) have identified the 
same transition and its phase boundary has been obtained both experimentally 
and theoretically \cite{ieee}.

\vskip 0.1in
{\centering \resizebox*{4.0in}{1.3in}{\rotatebox{0}{\includegraphics{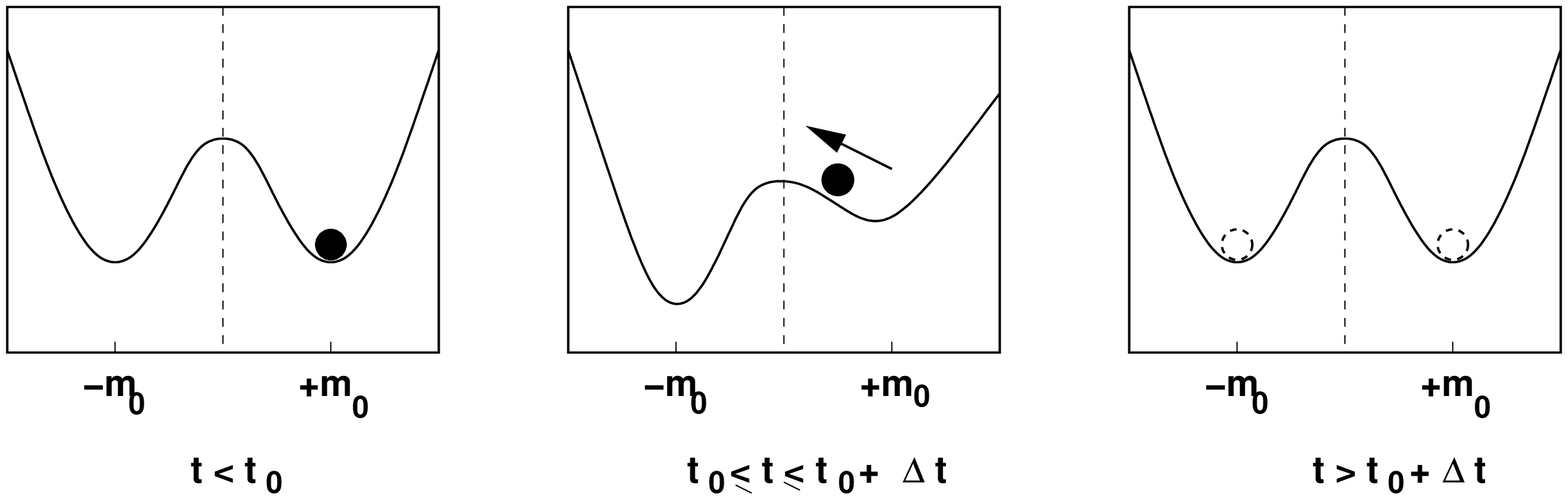}}} \par}
\vspace{0.3cm}
\vskip 0.1in
\noindent {\footnotesize FIG. 1. {
Schematic diagram showing the free energy landscape against the magnetization 
at different times. For $t<t_{0}$ the landscape is symmetric with respect to 
$m=0$ and the system is in $+m_{0}$ state as shown by $\bullet$ . For 
$t_{0} \leq t \leq t_{0} + \Delta t$, the landscape acquires a global minimum 
at $m=-m_{0}$ due to the presence of the external field and the system starts 
moving towards it. Beyond $t=t_{0}+\Delta t$, the landscape again becomes 
symmetric and the system eventually stabilizes in either of the 
equivalent minima as shown by $\circ$.
}}{\footnotesize \par}
\vskip 0.1in

\section {Model and the Transition}
The model studied here is the Ising model with nearest-neighbour interaction
under a time-dependent external magnetic field. This is described
by the Hamiltonian:
\begin{equation}
\label{ham}
H=-\sum _{\{ij\}}J_{ij}S_{i}S_{j}-h(t)\sum _{i}S_{i},
\end{equation}

\noindent where \( J_{ij} \) is the cooperative interaction between
the spins at site \( i \) and \( j \) respectively, and each nearest-neighbour
pair is denoted by \{...\}. We consider a square lattice. The static
critical temperature is \( T_{c}^{0}=2/\ln (1+\sqrt{2})\simeq 2.269... \) (in units
of \( J/K_{B} \)). At \( T<T_{c}^{0} \), an external field pulse
is applied, after the system is brought to equilibrium characterised
by an equilibrium magnetisation \( m_{0}(T) \). The spatially uniform
field has a time-dependence as follows:\begin{equation}
\label{hpdef}
h(t)=\left\{ \begin{array}{cc}
-h_{p} & t_{0}\leq t\leq t_{0}+\Delta t\\
0 & \rm otherwise.
\end{array}\right. 
\end{equation}

\vskip 0.5in
{\centering \resizebox*{8cm}{6.5cm}{\rotatebox{0}{\includegraphics{ptrfig2.eps}}} \par}
\vspace{0.3cm}
\vskip 0.1in
\noindent {\footnotesize FIG. 2. {Typical time variation of the response 
magnetisation \( m(t) \) for two different field pulses \( h(t) \) with same 
\( \Delta t \) for an Ising system at a fixed temperature \( T \).  The 
magnetisation-reversal here occurs due to increased pulse strength, keeping 
their width \( \Delta t \) same. The transition can also be brought about by 
increasing \( \Delta t \), keeping \( h_{p} \) fixed. The inset indicates the 
typical phase boundaries (where the field withdrawal-time magnetisation 
\( m_{w}=0 \)) for two different temperatures (sequential updating; note that 
for random updating the phase boundaries shift upwards).}}{\footnotesize \par}
\vskip 0.1in

Typical time-dependent (response) magnetisation \( m(t) \) (= \( <S_{i}>, \)
where $<\ldots>$ denotes the thermodynamic `ensemble' average) of the system
under different magnetic field \( h(t) \) are indicated in the Fig. 2.
The time \( t_{0} \) at which the pulse is applied is chosen such
that the system reaches its equilibrium at \( T \) ($<$ \( T_{c}^{0} \)
). As soon as the field is applied, the magnetisation \( m(t) \)
starts decreasing, continues until time \( t+\Delta t \) when the
field is withdrawn. The system relaxes eventually to one of the two
equlibrium states (with magnetisation \( -m_{0} \) or \( +m_{0} \)).
At a particular temperature \( T \), for appropriate combinations
of \( h_{p} \) and \( \Delta t, \) a magnetisation-reversal transition
occurs, when the magnetisation of the system switches from one state
of equilibrium magnetisation \( m_{0} \) to the other with magnetisation
\( -m_{0} \). This reversal phenomena at \( T<T_{c}^{0} \) is simple
and well studied for \( \Delta t\rightarrow \infty  \) for any non-zero
\( h_{p} \). We study here the dynamics for finite \( \Delta t \)
values. It appears that generally \( h_{p} \rightarrow \infty \) as
\( \Delta t\rightarrow 0 \) and \( h_{p} \rightarrow 0 \) as 
 \( \Delta t\rightarrow \infty  \) for any such dynamic magnetisation-reversal
transition phase boundary at any temperature \( T \) ($<$ \( T_{c}^{0} \)).
In fact, a simple application of the domain nucleation theory gives 
\(h_p \rm{ln} \Delta t\) = constant along the phase boundary, where the constant
changes by a factor \(1/(d+1)\), where \(d\) denotes the lattice dimension,
 as the boundary changes from single to multi-domain region \cite{pul1}.

\section {Mean-field study}
Consider a system of \( N \) Ising spins in contact with a heat bath evolving 
under Glauber single spin flip dynamics.
The master equation can be written as \cite{sk68}

\begin{eqnarray*}
\frac{d}{dt}P\left( S_{1},\cdots ,S_{N};\, t\right)  & = & -\sum _{j}W_{j}(S_{j})P\left( S_{1},\ldots ,S_{N};\, t\right) + \sum _{j}W_{j}(-S_{j})P\left( S_{1},\ldots ,-S_{j},\ldots ,S_{N};\, t\right) ,
\label{master}
\end{eqnarray*}
where \( P\left( S_{1},\cdots ,S_{N};\, t\right)  \) is the probability to
find the spins in the configuration \( \left( S_{1},\cdots ,S_{N}\right)  \)
at time \( t \) and \( W_{j}(S_{j}) \) is the probability of flipping of the
\( j \)th spin. Satisfying the condition of detailed balance one can write
the transition probability as

\begin{eqnarray*}
& W_{j}(S_{j}) = \frac{1}{2\lambda }\left[ 1-S_{j}\tanh \left( \frac{\sum _{i}J_{ij}S_{i}(t)+h_{j}}{T}\right) \right] ,&
\label{tran-prob}
\end{eqnarray*}
where \( \lambda  \) is a temperature dependent constant. Defining the spin
expectation value as

\begin{eqnarray*}
 & m_{i} = \left\langle S_{i}\right\rangle = \sum_{S} S_{i}P\left( S_{1},\ldots ,S_{N};\, t\right) ,& 
\label{m=<s>}
\end{eqnarray*}
where the summation is carried over all possible spin configurations, one can
write

\begin{equation}
\label{d<s>/dt}
\lambda \frac{dm_{i}}{dt}=-m_{i}+\left\langle \tanh \left( \frac{\sum _{j}J_{ij}S_{j}+h_{i}}{T}\right) \right\rangle .
\end{equation}
Using the mean field approximation linearisation (in $m$), (\ref{d<s>/dt}) 
can be written after a Fourier transform,

\begin{equation}
\label{lin dmq/dt}
\frac{dm_{q}(t)}{dt}=\lambda ^{-1}\left[ \left( K(q)-1\right) m_{q}(t)+\frac{h_{q}(t)}{T}\right] ,
\end{equation}
 where \( K(q)=J(q)/T \). When we are concerned only with the homogeneous magnetization,
we consider the \( q=0 \) mode of the equation and writing \( m_{q=0}=m \)
and \( h_{q=0}=h \), we get

\begin{equation}
\label{lin dm/dt}
\frac{dm}{dt}=\lambda ^{-1}\left[ \left( K(0)-1\right) m(t)+\frac{h(t)}{T}\right] .
\end{equation}
 In the mean field approximation \( K(0)=T_{c}^{MF}/T \) with \( T_{c}^{MF}=J(0) \)
and for small \( q \), \( K(q)\simeq K(0)\left( 1-q^{2}\right)  \). Differentiating
(\ref{lin dmq/dt}) with respect to the external field, we get the rate equation
for the dynamic susceptibility \( \chi _{q}(t) \) as

\begin{equation}
\label{lin dx/dt}
\frac{d\chi _{q}(t)}{dt}=\lambda ^{-1}\left[ \left( K(q)-1\right) \chi _{q}(t)+\frac{1}{T}\right] .
\end{equation}

We divide
the entire time zone in three different regimes : (I) \( 0<t<t_{0} \), where
\( h(t)=0 \) (II) \( t_{0}\leq t\leq t_{0}+\Delta t \), where \( h(t)=-h_{p} \)
and (III) \( t_{0}+\Delta t<t<\infty  \), where \( h(t)=0 \) again. We note
that (\ref{lin dm/dt}) can be readily solved separately for the three regions
as the boundary conditions are exactly known. In region I, \( dm/dt=0 \) and
the solution of the linearized (\ref{lin dm/dt}) becomes trivial. We, therefore,
use the solution of (\ref{lin dmq/dt}) in region I (\( m_{0}=\tanh \left( m_{0}T_{c}^{MF}/T\right)  \))
as the initial value of \( m \) for region II. Integrating (\ref{lin dm/dt})
in region II, we then get

\begin{equation}
\label{m(t) in II}
m(t)=\frac{h_{p}}{\Delta T}+\left( m_{0}-\frac{h_{p}}{\Delta T}\right) \exp \left[ b\Delta T\left( t-t_{0}\right) \right] ,
\end{equation}
 where \( b=1/\lambda T \) and \( \Delta T=T_{c}^{MF}-T \). We also note 
that in order to justify the validity of the linearization of (\ref{lin dmq/dt})
one must keep the factor inside the exponential of (\ref{m(t) in II}) small.
This restricts the linear theory to be valid close to \( T_{c}^{MF} \)
and for small  \( \Delta t \). Writing \( m_{w}\equiv m(t_{0}+\Delta t) \),
we get from (\ref{m(t) in II})

\begin{equation}
\label{mw}
m_{w}=\frac{h_{p}}{\Delta T}+\left( m_{0}-\frac{h_{p}}{\Delta T}\right) e^{b\Delta T\Delta t}.
\end{equation}
 It is to be noted here that in absence of fluctuations, the sign of \( m_{w}(h_{p},\Delta t) \)
solely decides which of the final two equilibrium states will be chosen by the
system after the withdrawal of the pulse. At \( t=t_{0}+\Delta t \), if \( m_{w}>0 \),
the system gets back to \( +m_{0} \) state and if \( m_{w}<0 \), magnetization-reversal
transition occurs and the system eventually chooses the \( -m_{0} \) state
(see figure 2). Thus setting \( m_{w}=0 \), we obtain the threshold value of
the pulse strength at the mean field phase boundary for this dynamic phase transition.
At any \( T \), combinations of \( h_{p} \) and \( \Delta t \) below the
phase boundary cannot induce the magnetization-reversal transition, while those
above it can induce the transition. From (\ref{mw}) therefore we can write
the equation of the mean field phase boundary for the magnetization-reversal
transition as

\begin{equation}
\label{phase-bound}
h_{p}^{c}(\Delta t,T)=\frac{\Delta Tm_{0}}{1-e^{-b\Delta T\Delta t}}.
\end{equation}

\begin{center}
\resizebox*{3.1in}{2.4in}{\rotatebox{270}{\includegraphics{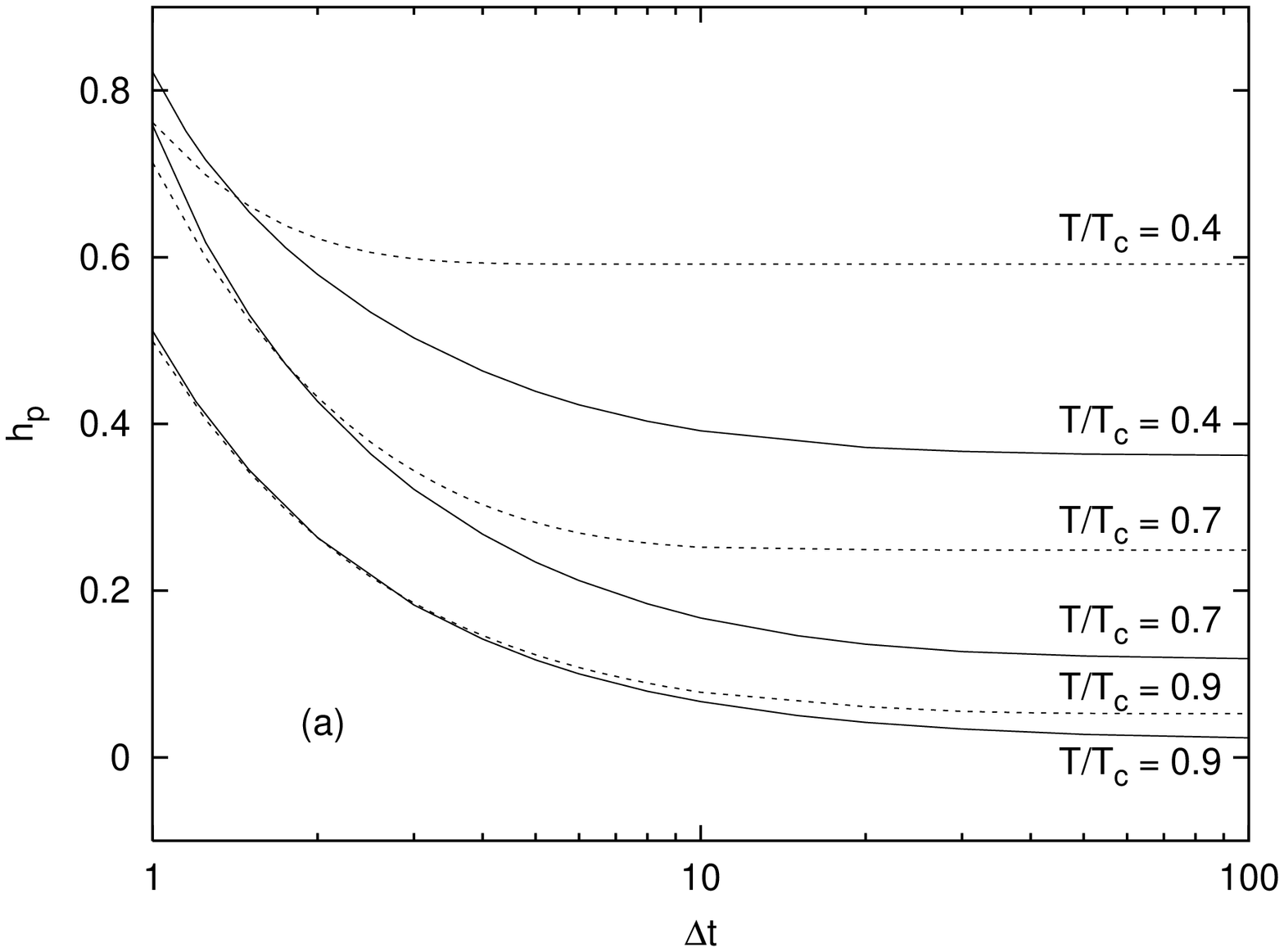}}}
\resizebox*{3.1in}{2.4in}{\rotatebox{270}{\includegraphics{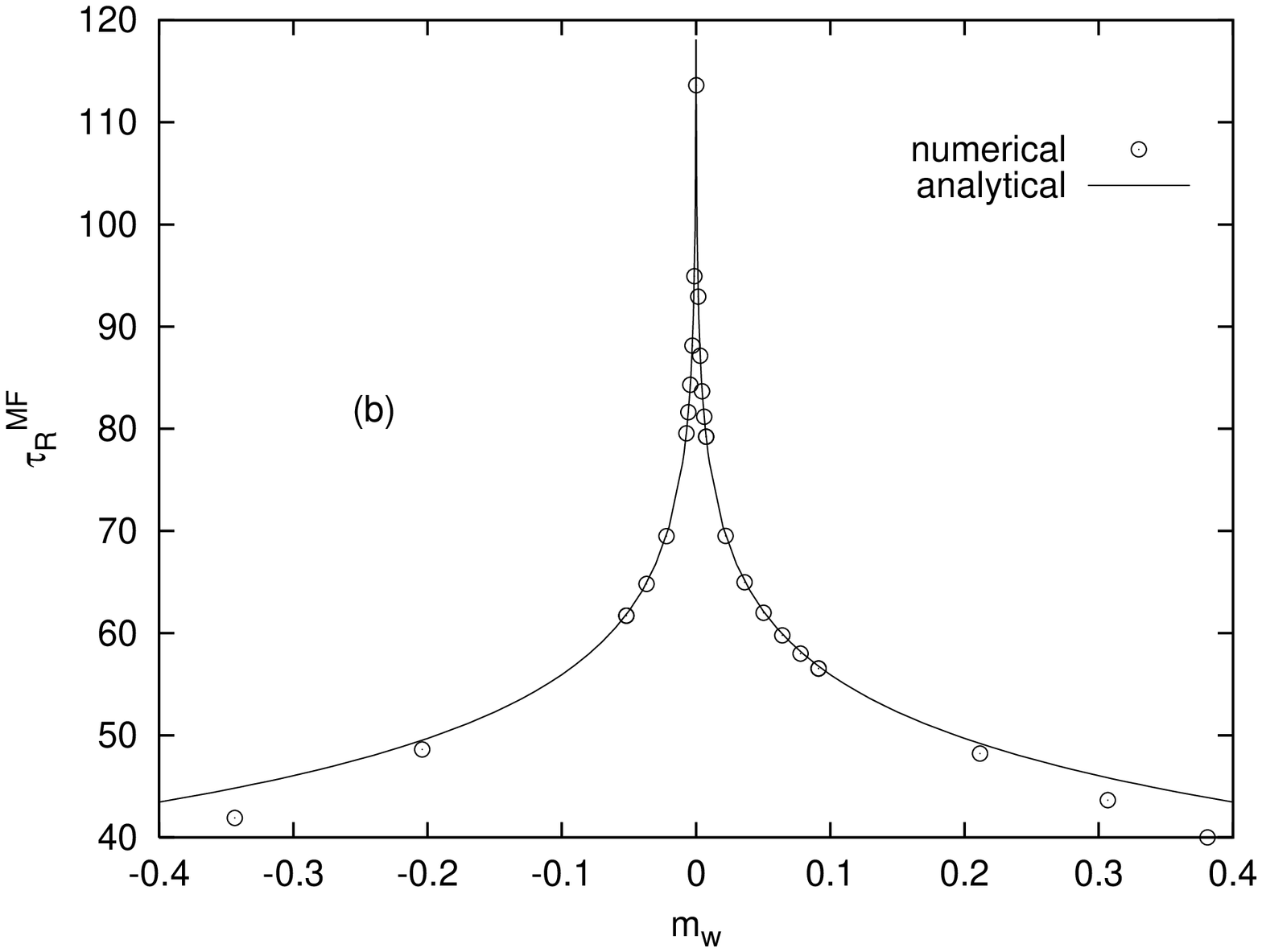}}}
\end{center}
\vspace{0.3cm}
\vskip 0.1in
\noindent {\footnotesize FIG. 3. {
(a) MF phase boundaries for three different temperatures.
The solid line is obtained from numerical solution of (\ref{d<s>/dt}) in MF 
case and the dotted lines give the corresponding analytical estimates in the 
linear limit (from \cite{mish1-1}). 
(b) Logarithmic divergence of $\tau_R^{MF}$ across the phase 
boundary for $T/T_c=0.9$. The data points (circles) for $\tau_R^{MF}$ are 
obtained from the solution of (\ref{lin dmq/dt}) (with $q=0$, $|m(t)-m_w| 
< 10^{-4}$) and the solid line corresponds to (\ref{tau})
is the solution of the linearized MF equation (from \cite{mish1-1}).
}}{\footnotesize \par}
\vskip 0.1in

Figure 3a shows phase boundaries at different \( T \) obtained from (\ref{phase-bound})
and compares those to the phase boundaries obtained from the numerical solution
of the full dynamical equation (\ref{lin dmq/dt}). The phase boundaries obtained
under linear approximation match quite well with those obtained numerically
for small values of \( \Delta t \) and at temperatures close to \( T_{c}^{MF} \),
which is the domain of validity of the linearized theory as discussed before.
In region III, we again have \( h(t)=0 \) and solution of (\ref{lin dm/dt})
leads to

\begin{equation}
\label{m(t) in III}
m(t)=m_{w}\exp \left[ b\Delta T\left\{ t-\left( t_{0}+\Delta t\right) \right\} \right] .
\end{equation}
 We define the relaxation time \( \tau _{R}^{MF} \), measured from \( t=t_{0}+\Delta t \),
as the time required to reach the final equilibrium state characterized by magnetization
\( \pm m_{0} \) in region III (see figure 1). From above expression therefore
we can write

\begin{eqnarray}
\tau _{R}^{MF} & = & \frac{1}{b\Delta T}\ln \left( \frac{m_{0}}{\left| m_{w}\right| }\right) \nonumber \\
 & \sim  & -\left( \frac{T}{T_{c}^{MF}-T}\right) \ln \left| m_{w}\right| .\label{tau}
\end{eqnarray}
A point to note is that \( m(t) \) in (\ref{m(t) in III}) grows exponentially
with \( t \) and therefore in order to confine ourselves to the linear regime
of \( m(t) \), \( m_{0} \) must be small (\( T \) close to \( T_{c}^{MF} \))
and \( t\leq \tau _{R}^{MF} \). The factor \( \left( T_{c}^{MF}-T\right) ^{-1} \)
gives the usual critical slowing down for the static transition at \( T=T_{c}^{MF} \).
However, even for \( T\ll T_{c}^{MF} \), \( \tau _{R}^{MF} \) diverges at
the magnetization-reversal phase boundary where \( m_{w} \) vanishes. Figure
3b shows the divergence of \( \tau ^{MF}_{R} \) against \( m_{w} \) as obtained
from the numerical solution of the full mean field equation of motion 
(\ref{lin dmq/dt}) and compares it with that obtained from (\ref{tau}).

Solution of \( \chi _{q}(t) \) is more difficult as all the boundary conditions
are not directly known. However, \( \chi _{q}(t) \) can be expressed in terms
of \( m(t) \) and the solution of the resulting equation will then have the
\( t \) dependence coming through \( m(t) \), which we have solved already.
Dividing (\ref{lin dx/dt}) by (\ref{lin dm/dt}) we get

\begin{equation}
\label{dx/dm}
\frac{d\chi _{q}(t)}{dm}=\frac{\chi _{q}(t)+ \frac{1}{T(K(q)-1)}}{m(t)\left(\frac{K(0)-1}{K(q)-1}\right ) + \frac{h(t)}{K(q)-1}},
\end{equation}
 which can be rewritten in the linear limit as

\begin{equation}
\label{lin dx/dm}
\frac{d\chi _{q}}{\chi _{q}+\Gamma }=a_{q}\frac{dm}{m+h(t)/\Delta t},
\end{equation}
 where \( \Gamma =1/T\left( K(q)-1\right)  \) and \( a_{q}=\left( K(q)-1\right) /\left( K(0)-1\right) \simeq 1-q^{2}/\Delta T \)
for small \( q \).

In region II, solution of (\ref{lin dx/dm}) can be written as

\begin{equation}
\label{x(t) in II}
\chi _{q}(t)=-\Gamma +\left( \chi _{q}^{s}+\Gamma \right) \left[ \frac{m(t)-h_{p}/\Delta T}{m_{0}-h_{p}/\Delta T}\right] ^{a_{q}},
\end{equation}
 where \( \chi _{q}^{s} \) is the equilibrium value of susceptibility in region
I. Solving (\ref{x(t) in II}) in region III with the initial boundary condition
\( m\left( t_{0}+\Delta t\right) =m_{w} \), we get

\begin{eqnarray}
\chi _{q}(t) & = & -\Gamma +\left( \chi _{q}\left( t_{0}+\Delta t\right) +\Gamma \right) \left( \frac{m(t)}{m_{w}}\right) ^{a_{q}}\nonumber \\
 & = & -\Gamma +\left( \chi _{q}^{s}+\Gamma \right) \left( \frac{m(t)}{m_{w}}\right) ^{a_{q}}e^{b\Delta T\Delta ta_{q}},\label{x(t) in III}
\end{eqnarray}
 where use has been made of (\ref{x(t) in II}) and (\ref{mw}). The dominating
\( q \) dependence in \( \chi _{q}(t) \) is coming from \( \left( 1/m_{w}\right) ^{a_{q}} \)
when \( m_{w}\rightarrow 0 \) as one approaches the phase boundary. The singular
part of the dynamic susceptibility can then be written as

\begin{equation}
\label{x(q) with xi}
\chi _{q}(t)=\left( \chi _{q}^{s}+\Gamma \right) \exp \left[ -q^{2}\left( \xi ^{MF}\right) ^{2}\right] ,
\end{equation}
 where for small values of \( m_{w} \) the correlation length \( \xi ^{MF} \)
is given by \cite{mish1,mish1-1}

\begin{equation}
\label{xi}
\xi ^{MF}\equiv \xi ^{MF}\left( m_{w}\right) =\left[ \frac{T_{c}}{\Delta T}\ln \left( \frac{1}{\left| m_{w}\right| }\right) \right] ^{\frac{1}{2}}.
\end{equation}
 Thus the length scale also diverges at the magnetization-reversal phase boundary
and this can be demonstrated even using the linearized mean field equation of
motion. Equations (\ref{tau}) and (\ref{xi}) can now be used to establish
the following relation between the diverging time and length scales :

\begin{equation}
\label{tau-xi}
\tau _{R}^{MF}\sim \frac{T}{T_{c}}\left( \xi ^{MF}\right) ^{2},
\end{equation}
 which leads to a dynamical critical exponent \( z=2. \) 
One can have a numerical estimate of \( \xi ^{MF} \) by solving 
(\ref{x(q) with xi}) for different values of \( q \). Figure 4(b) shows plots of 
\( \chi _{q}(t) \) against \( m_{w} \) for different values of \( q \). 
It may be noted that these divergences in \( \tau _{R}^{MF} \) and 
\( \xi ^{MF} \) are shown to occur for any \( T<T_{C}^{MF} \), and these 
dynamic relaxation time and correlation length defined for the 
magnetization-reversal transition exist only for \( T<T_{c}^{MF} \).

It may further be noted from (\ref{x(q) with xi}) that \( \chi _{q}(t)\rightarrow 0 \)
as \( \xi ^{MF}\rightarrow \infty  \), thereby producing a minimum of \( \chi _{q} \)
at the phase boundary. The absence of any divergence in the susceptibility is
due to the fact that at \( t=t_{0}+\Delta t \), there remains no contribution
of \( m_{w} \) in \( \chi _{q}(t) \) as is evident from (\ref{x(t) in III}).
However, numerical solution of (\ref{dx/dm}) for \( q=0 \) mode shows a clear
singularity in the homogeneous susceptibility \( \chi _{0} \) at the 
magnetization-reversal phase boundary ($m_{w}=0$), as depicted in figure 4a. 

{\centering \resizebox*{3.0in}{2.3in}{\rotatebox{270}{\includegraphics{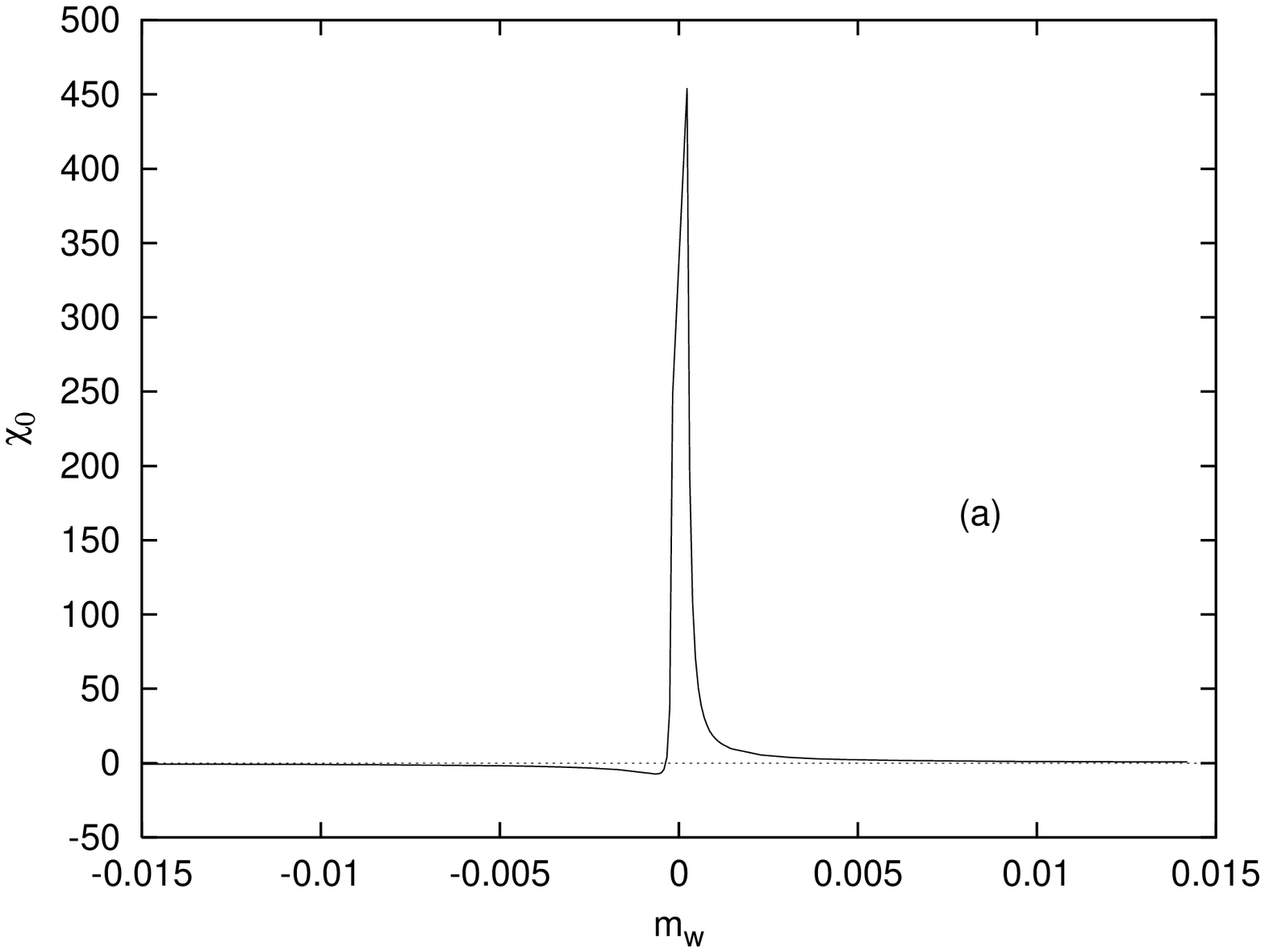}}}}
{\centering \resizebox*{3.0in}{2.3in}{\rotatebox{270}{\includegraphics{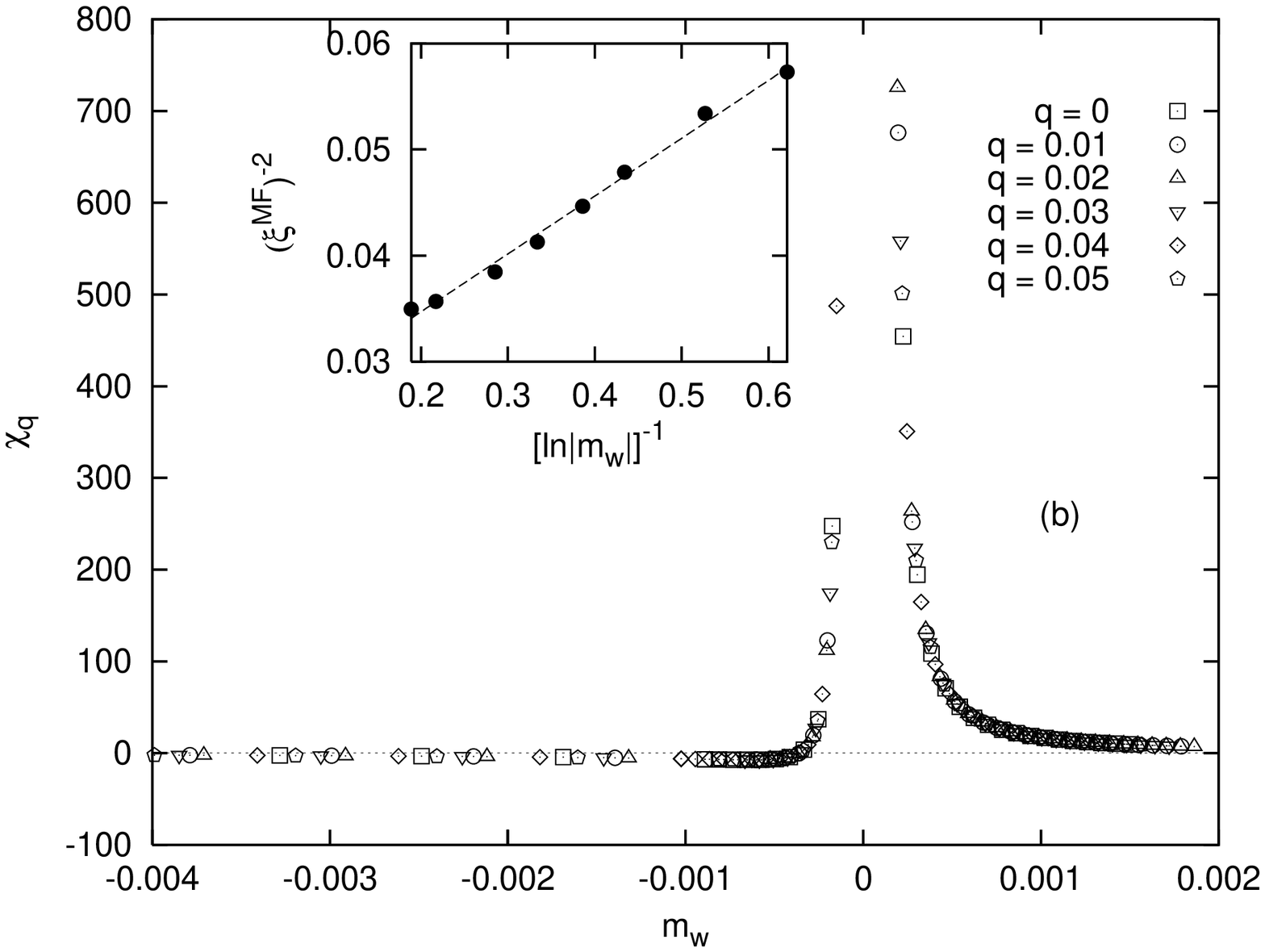}}} \par}
\vspace{0.3cm}
\vskip 0.1in
\noindent {\footnotesize FIG. 4. {
(a) Divergence of \( \chi _{q=0} \) across the phase boundary obtained from the 
numerical solution of (\ref{dx/dm}). (b) Plot of \( \chi _{q} \) against 
\( m_{w} \) for different values of \( q \). The inset shows the linear 
variation of \( \left( \xi ^{MF}\right) ^{-2} \)
against \( \left[ \ln \left| m_{w}\right| \right] ^{-1} \). The data points
for \( \xi ^{MF} \) in the inset are obtained from the slope of the best fitted
straight lines through a plot of \( \ln \chi _{q} \) against \( q^{2} \) for
different values of \( m_{w} \) (from \cite{mish1-1}).
}}{\footnotesize \par}
\vskip 0.1in

\section {Monte-Carlo study and the results}

Here the Monte-Carlo study has been carried out in 2D and 3D (square and cubic
lattices) with periodic boundary conditions. Spins are updated
following Glauber dynamics. The updating rule employed here are both
sequential as well as random. In sequential updating rule one Monte-Carlo
step consist of a complete scan of the lattice in a sequential manner;
while in random updating a Monte-Carlo step is defined by \( N \)
($= L^{d}$) random updates on the lattice, where $N$ is the total number 
of spins in a lattice of linear size $L$ in $d$ dimensions. Studies
have been carried out at temperatures below the static critical temperature
(e.g, $T_{c}^{0}\simeq 2.27$ for 2D). The system is allowed to evolve from an 
initial state of perfect order to its equilibrium state at temperature \( T \). 
The time \( t_{0} \) is chosen to be
much larger than the static relaxation time at that \( T \), so that
the system reaches an equilibrium state with magnetization \( +m_{0}(T) \)
before the external magnetic field is applied at time \( t=t_{0} \).
The field pulse of strength \( -h_{p} \) is applied for duration
\( \Delta t \) (measured in Monte Carlo steps or MCS). The magnetisation
starts decreasing from its equilibrium value \( m_{0} \). The average
value of the magnetisation \( m_{w} \) at the time of withdrawal
of pulse is noted. The phase boundary of this dynamic transition is
defined by appropriate combination of \( h_{p} \) and \( \Delta t \)
that produces the magnetisation reversal by making \( m(t_{0}+\Delta t)\equiv m_{w}=0 \)
from a value \( m(t_{0})=m_{0}, \) i.e, \( m_{w} \) changes sign
across the phase boundary. The phase boundary changes with \( T. \)
The behavior of different thermodynamic quantities are studied across
the phase boundary. These quantities are averaged over different initial 
configurations of the system. The fluctuations over the average value 
are also noted. 

We note the marked qualitative difference between the MF and MC phase 
boundaries (see Fig. 3 and Fig. 5). In the MF case, even for 
$\Delta t \rightarrow \infty$,
due to absence of fluctuations, $h_p$ must be greater than the non-zero 
coercive field in order that the transition takes place and the phase
boundary becomes flat for large $\Delta t$. In MC study, due to presence
of fluctuations, even a pulse of infinitesimal strength, if applied for 
very long time, can induce the transition. This is evident from the
asymptotic nature of the phase boundaries for large $\Delta t$.

\begin{center}
\resizebox*{3.1in}{2.3in}{\rotatebox{0}{\includegraphics{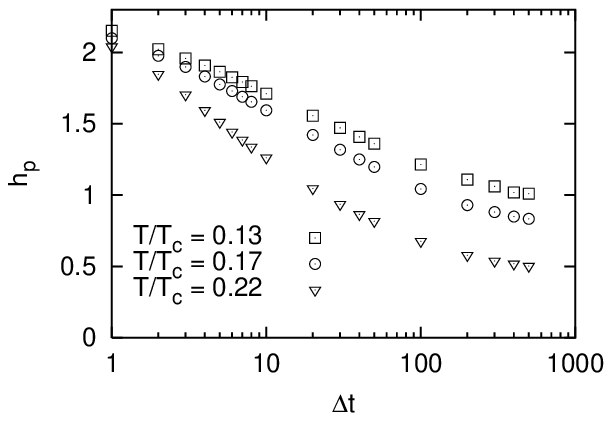}}}
\resizebox*{3.1in}{2.3in}{\rotatebox{0}{\includegraphics{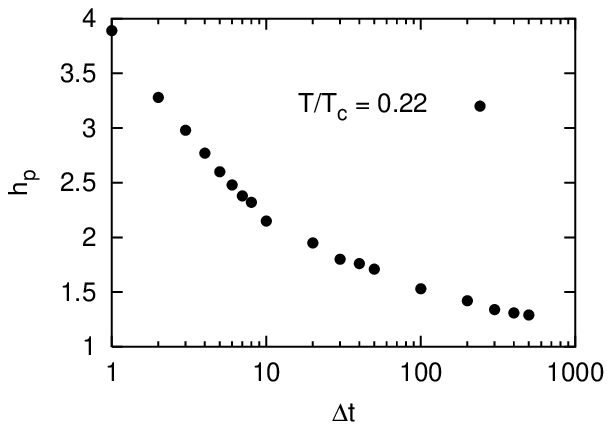}}}
\end{center}
\vspace{0.3cm}
\vskip 0.1in
\noindent {\footnotesize FIG. 5. {
Phase boundaries from Monte Carlo studies on square lattice with $L=100$
for sequential updating (open symbols) \cite{mish1-1} and random updating 
(filled symbol).
}}{\footnotesize \par}
\vskip 0.1in

To understand the nature of the MC phase diagram of the magnetization-reversal 
transition we consult the classical theory of nucleation (CNT). 
A typical of a ferromagnet, below its static critical temperature
\( T_{c}^{0} \), consists of droplets or domains of spins oriented in the same
direction, in a sea of oppositely oriented spins. According to CNT, the 
equilibrium number of droplets consisting of \( s \) spins is given by 
\( n_{s}=N\exp \left( -\epsilon _{s}/T\right)  \),
where \( \epsilon _{s} \) is the free energy of formation of a droplet 
containing \( s \) spins and \( N \) being the normalization constant. 
In presence of a negative external magnetic field \( h \), the free energy 
can be written as \( \epsilon _{s}=-2hs+\sigma s^{(d-1)/d} \),
where the droplet is assumed to be spherical and \( \sigma (T) \) 
is the temperature dependent surface tension. Droplets of size greater than
a critical value \( s_{c} \) are favoured to grow, where 
\( s_{c}=\left[ \sigma (d-1)/(2d\left| h\right| )\right] ^{d} \)
is obtained by maximizing \( \epsilon _{s} \). The number of supercritical
droplets is therefore given by 
\(n_{s_{c}}=N\exp \left[ -\Lambda_{d}\sigma^{d}\left| h\right|^{1-d}/T\right]\),
where \( \Lambda_{d} \) is a constant depending on dimension only. In the SD 
regime, where a single supercritical droplet grows to eat up the whole system,
the nucleation time goes inversely as the nucleation rate \( I \).
According to the Becker-D\"{o}ring theory, \( I \) is proportional
to \( n_{s_{c}} \) and therefore one can write
\[
\tau^{SD}_{N}\propto I^{-1}\propto \exp \left[ \frac{\Lambda _{d}\sigma ^{d}}{T\left| h\right| ^{d-1}}\right] .\]
 However, in the MD regime the nucleation mechanism is different and in this
regime many supercritical droplets grow simultaneously and eventually coalesce
to create a system spanning droplet. The radius \( s_{c}^{1/d} \) of a supercritical
droplet grows linearly with time \( t \) and thus \( s_{c}\propto t^{d} \).
For a steady rate of nucleation, the rate of change of magnetization is \( It^{d} \).
For a finite change \( \Delta m \) of the magnetization during the nucleation
time \( \tau _{N}^{MD} \), one can write
\[
\Delta m\propto \int _{0}^{\tau ^{MD}_{N}}It^{d}dt=I\left( \tau _{N}^{MD}\right)^{d+1}.\]
 Therefore, in the MD regime one can write \cite{rtms94}\cite{as98}
\begin{equation}
\label{taumd}
\tau _{N}^{MD}\propto I^{-1/(d+1)}\propto \exp \left[ \frac{\Lambda _{d}\sigma ^{d}}{T(d+1)\left| h\right| ^{d-1}}\right] .
\end{equation}

 During \( t_{0}\leq t\leq t_{0}+\Delta t \), when the external field is 
`on', the only relevant time scale in the system is the nucleation time. The 
magnetization reversal phase boundary gives the threshold value \( h_{p}^{c} \)
of the pulse strength which, within time \( \Delta t \), brings the system
from an equilibrium state with magnetization \( +m_{0} \) to a non-equilibrium
state with magnetization \( m_{w}=0_{-} \), so that eventually the system evolves
to the equilibrium state with magnetization \( -m_{0} \). The field driven
nucleation mechanism takes place for \( t_{0}\leq t\leq t_{0}+\Delta t \) and
therefore equating the above nucleation times with \( \Delta t \), one gets
for the magnetization-reversal phase boundary
\begin{equation}
\label{logdt-hp}
\begin{array}{cccccl}
\ln \left( \Delta t\right)  & = & c_{1} & + & C\left[ h_{p}^{c}\right] ^{1-d}, & \textrm{in the SD regime}\\
 & = & c_{2} & + & C\left[ h_{p}^{c}\right] ^{1-d}/(d+1), & \textrm{in the MD regime}
\end{array}
\end{equation}
\noindent where \( C=\Lambda _{d}\sigma ^{d}/T \) and \( c_{1} \), \( c_{2} \)
are constants. Therefore a plot of \( \ln (\Delta t) \) against \( \left[ h_{p}^{c}\right] ^{d-1} \)
would show two different slopes corresponding to the two regimes \cite{mc00}.
Figure 6 shows these plots and it indeed have two distinct slopes for both \( d=2 \)
(figure 6(a)) and \( d=3 \) (figure 6(c)) at sufficiently high temperatures,
where both the regimes are present. The ratio \( R \) of the slopes corresponding
to the two regimes has got values close to \( 3 \) for \( d=2 \) and close
to \( 4 \) for \( d=3 \), as suggested by (\ref{logdt-hp}). The value of
\( h_{DSP} \) is obtained from the point of intersection of the straight lines
fitted to the two regimes. At lower temperatures, however, the MD region is
absent and the phase diagram here is marked by a single slope as shown in 
figures 6(b) and 6(d).

The relaxation behaviour follows :
\begin{equation}
\label{MCtau}
\tau _{R}\sim \kappa (T,L)e^{-\mu (T)\left| m_{w}\right| },
\end{equation}
 where \( \kappa (T,L) \) is a constant depending on temperature and system
size and \( \mu (T) \) is a constant depending on temperature only. It may
be noted from (\ref{MCtau}) that \( \tau \rightarrow \kappa (T,L) \) as 
\( m_{w}\rightarrow 0 \). Therefore the true divergence at the phase boundary 
(where \( m_{w}=0 \)) of the relaxation time depends on the nature of 
\( \kappa (T,L) \). \( \kappa (T,L) \) grows sharply with system size.
The relaxation time \( \tau _{R} \) therefore diverges in the thermodynamic
limit (\( L\rightarrow \infty  \)) through the constant \( \kappa  \). It
may be noted that this divergence of \( \tau _{R} \) at the dynamic 
magnetization-reversal phase boundary occurs even at temperatures far below 
the static critical temperature \( T_{c}^{0} \).

{\centering \resizebox*{3.7in}{2.6in}{\rotatebox{0}{\includegraphics{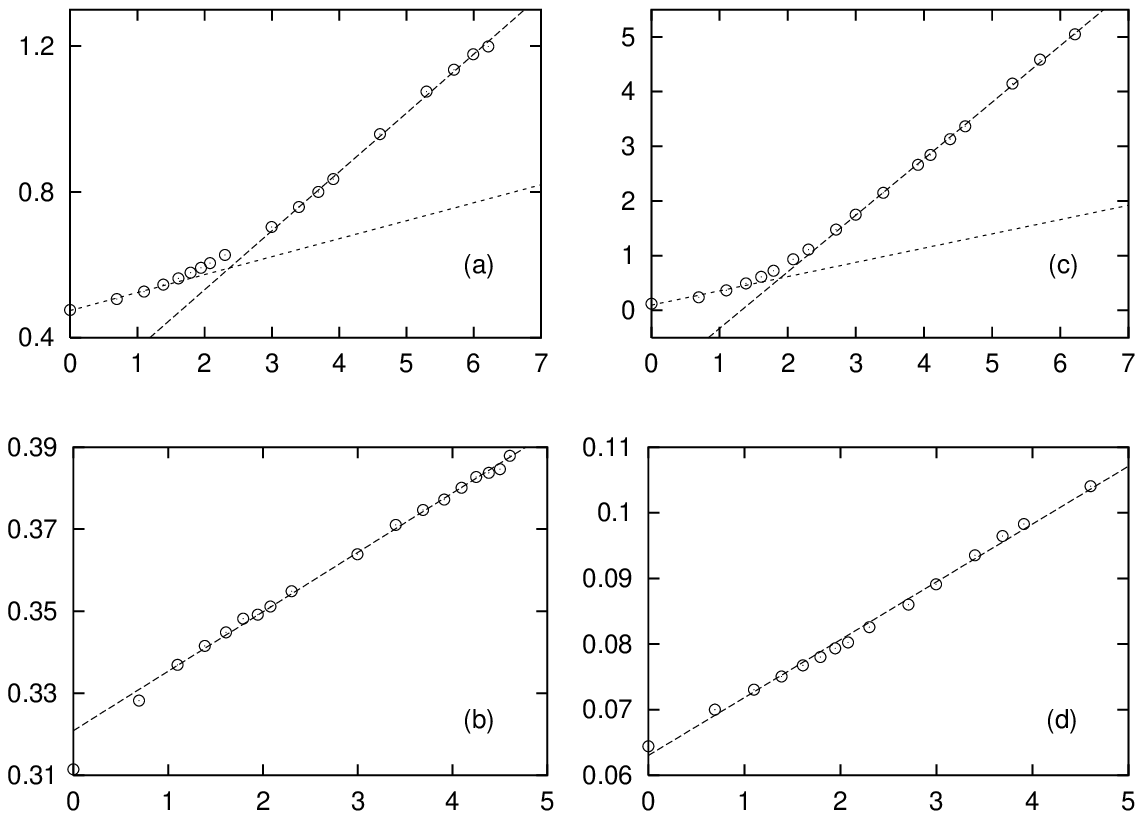}}} \par}
\vspace{0.3cm}
\vskip 0.1in
\noindent {\footnotesize FIG. 6. {
Plots of \( \ln \Delta t \) against \( \left( h_{p}\right) ^{1-d} \)
along the MC phase boundary. (a) \( T/T_{c}=0.31 \) and (b) \( T/T_{c}=0.09 \)
for square lattice and (c) \( T/T_{c}=0.67 \) and (d) \( T/T_{c}=0.11 \) for
simple cubic lattice. The slope ratio \( R\simeq 3.27 \) in (a) and 
\( \simeq 3.97 \) in (c) (from \cite{mish1-1}).
}}{\footnotesize \par}
\vskip 0.1in

{\centering \resizebox*{4.0in}{1.6in}{\rotatebox{0}{\includegraphics{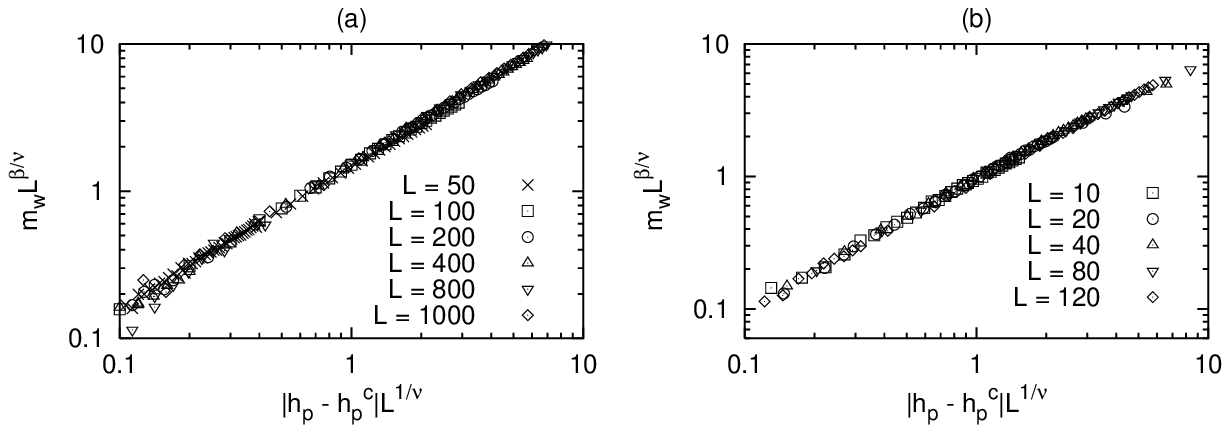}}} \par}
\vspace{0.3cm}
\vskip 0.1in
\noindent {\footnotesize FIG. 7. {
Finite size scaling fits : (a) for \( d=2 \) at \( T/T_{c}=0.88 \) and 
(b) for \( d=3 \) at \( T/T_{c}=0.67 \) (from \cite{mish1-1}).
}}{\footnotesize \par}
\vskip 0.1in

In the region where the transition is continuous, the scaling arguements are
expected to hold. We assume power law behaviour in this regime both for \( m_{w} \)
\begin{equation}
\label{powerlaw mw}
m_{w}\sim \left| h_{p}-h_{p}^{c}\left( \Delta t,T\right) \right| ^{\beta }
\end{equation}
 and for the correlation length
\begin{equation}
\label{powerlaw xi}
\xi \sim \left| h_{p}-h_{p}^{c}\left( \Delta t,T\right) \right| ^{-\nu }.
\end{equation}
For a finite size system, \( h_{p}^{c} \) is a function of the system size
\( L \). Assuming that at the phase boundary \( \xi  \) can at the most reach
a value equal to \( L \), one can write the finite size scaling form of \( m_{w} \)
as:

\begin{equation}
\label{fss}
m_{w}\sim L^{-\beta /\nu }f\left[ \left( h_{p}-h^{c}_{p}\left( \Delta t,T,L\right) \right) L^{1/\nu }\right] ,
\end{equation}
 where \( f(x)\sim x^{\beta /\nu } \) as \( x\rightarrow \infty  \). A plot
of \( m_{w}/L^{-\beta /\nu } \) against \( \left( h_{p}-h^{c}_{p}\left( \Delta t,T,L\right) \right) L^{1/\nu } \)
shows a nice collapse of the data for \( d=2 \) and for \( d=3 \) as shown in figure 7. The values of the critical exponents
obtained from the data collapse were \( \beta =0.85\pm 0.05 \) and \( \nu =1.5\pm 0.5 \)
in \( d=3 \) and \( \beta =1.00\pm 0.05 \) and \( \nu =2.0\pm 0.5 \) in \( d=2 \),
where \( h_{p}^{c}\left( \Delta t,T\right)  \) was obtained with an accuracy
\( O\left( 10^{-3}\right)  \). Attempts to fit similar data to the above
finite size scaling form obtained in the SD regime failed.

The behavior of the reduced fourth order cumulant \( U \) \cite{binheer} has 
also been studied near the magnetisation reversal transition boundary. 
This is defined as 
\begin{equation}
\label{bind}
U=1-\frac{\left\langle m_{w}^{4}\right\rangle }{3\left\langle m_{w}^{2}\right\rangle^{2}},
\end{equation}

\noindent where \( \left\langle m_{w}^{4}\right\rangle  \) is the
ensemble average of \( m_{w}^{4} \). \( \left\langle m_{w}^{2}\right\rangle  \)
is similarly defined. The cumulant \( U \) here behaves 
somewhat differently, compared to that in static and other transitions: Deep
inside the ordered phase \( m_{w}\simeq 1 \) and \( U\rightarrow 2/3. \)
For other (say, static) transitions the order parameter (\( m_{w} \))
goes to zero with a Gaussian fluctuation above the transition point,
 giving \( U\rightarrow 0 \) there. Here,
however, due to the presence of the pulsed field, \( \left| m_{w}\right|  \)
is non-zero on both sides of the magnetisation-reversal transition.
Hence \( U \) drops to zero at a point near the transition and grows again
after it.

\vskip 0.4in
\begin{center}
\resizebox{7.0cm}{5.0cm}{\rotatebox{0}{\includegraphics{ptrfig8a.eps}}}
\hspace{0.1in}
\resizebox{7.0cm}{5.0cm}{\rotatebox{0}{\includegraphics{ptrfig8b.eps}}} 
\vskip 0.4in
\resizebox{7.0cm}{5.0cm}{\rotatebox{0}{\includegraphics{ptrfig8c.eps}}}
\hspace{0.1in}
\resizebox{7.0cm}{5.0cm}{\rotatebox{0}{\includegraphics{ptrfig8d.eps}}} 
\end{center}
\vskip 0.1in
\noindent {\footnotesize FIG. 8. 
 Behavior of \( U \) near the transition,
driven by (a) \( T \) at a fixed value of \( h_p \) (=1.9) and \( \Delta t \)
(=\( 5 \)) with sequential updating, (b) \( h_p \) at a fixed value of \( T \)
(=0.5) and \( \Delta t \) (=\( 5 \)) with sequential updating, and 
(c) \( h_{p} \) at a fixed value of \( T \) (=0.5) and \( \Delta t \)
(=\( 5 \)) with random updating, for different \( L \), averaged over 
1000 to 20000 initial configurations. The insets show the typical 
behavior of the magnetisation \( m_{w} \) at the time of withdrawal of the
field pulse by varying (a) \( T \) at a fixed \( h_p \) and \( \Delta t \),
for \( L=100\) and \(800 \), (b) \( h_p \) at a fixed
\( T \) and \( \Delta t \), for \( L=100\) and \( 400 \), (c) \( h_{p} \) at a fixed \( T \) and \( \Delta t \),
for \( L=50\) and \( 200 \); \( m_{w}=0 \) at the effective transition point.
(d) Finite size scaling study in this parameter range: the effective \( T_c \)
or \( h_{p}^c \) values (see the insets), where \( m_{w}=0 \), are plotted 
against \(L^{-1/\nu}\) with \(\nu^{-1}=0.7\). The values of the cumulant 
crossing points in (a), (b), (c) are taken to correspond the respective 
transition points for \(L \rightarrow \infty\) (from \cite{acbkcpul}).
}{\footnotesize \par}
\vskip 0.1cm

The universality class of the dynamic transition in Ising model under
oscillating field has been studied extensively by investigating \cite{korn1}
the critical point and the cumulant value \( U^{*} \) at the critical
point, where the cumulant curves cross for different system sizes
(\( L \)). In that case, of course, the variation of \( U \) (at
any fixed \( L \)) is similar to that in the static Ising transitions
(\( U=2/3 \) well inside the ordered phase and \( U\rightarrow 0 \)
well within the disordered phase). In fact, \( U^{*} \) value in this 
oscillatory field case was found to be the same as that in the static case, 
indicating the same universality class \cite{korn1}. We have studied the 
behavior of $U$ in 2D. We observe different behavior in the field pulse 
induced magnetisation-reversal transition case.

We observed \cite{acbkcpul} two distinct behavior of the cumulant \( U \).
Typically, for low temperature and low pulse-duration region (see
the inset in Fig. 2) of the magnetisation-reversal phase boundary,
the cumulant crossing for different system sizes (\( L \)) occur
at \( U^{*}\simeq 0.42 \) to \( 0.46 \) (see Fig. 8). We
checked these results for both sequential and random updating. 
Specifically, for \( T=0.5 \) and \( \Delta t=5, \) (see Fig. 8c) for random 
updating we find the transition point value of \( h_{p}\simeq 2.6, \) to be 
smaller than the value (\( \simeq 1.9 \)) for sequential updating. However, 
the value of \( U^{*} \) at this transition point is very close to about 
\( 0.44 \). This indicates that updating rule does not affect the universality 
class (\( U^{*} \) value), as long as the proper region of the phase boundary 
is considered.
For relatively higher temperature and pulse-duration region of the
phase boundary, the crossing of \( U \) for different \( L \) values
occur for \( U^{*}\simeq 0_{+} \). This occurs both for sequential (Fig. 
9a, b) and random (Fig. 9c) updating. It may be noted that the phase
boundary changes with the updating rule, as the system relaxation
time (which matches with the pulse width at the phase boundary) is
different for sequential and random updating \cite{binheer}.

\vskip 0.1in
\vspace{0.3cm}
\begin{center}
\resizebox{7.0cm}{5.0cm}{\rotatebox{0}{\includegraphics{ptrfig9a.eps}}}
\resizebox{7.0cm}{5.0cm}{\rotatebox{0}{\includegraphics{ptrfig9b.eps}}}
\vskip 0.4in
\resizebox*{7.0cm}{5.0cm}{\rotatebox{0}{\includegraphics{ptrfig9c.eps}}}
\end{center}
\vskip 0.1in

\noindent {\footnotesize FIG. 9. 
 Behavior of \( U \) near the transition,
driven by (a) \( h_{p} \) at a fixed value of \( T \) (=2.0) and \( \Delta t \)
(=\( 5 \)) with sequential updating, (b) \( T \) at a fixed value of \( h_p \)
(=0.5) and \( \Delta t \) (=\( 10 \)) with sequential updating, and 
(c) \( h_{p} \) at a fixed value of \( T \) (=1.5) and \( \Delta t \)
(=\( 5 \)) with random updating,
for different \( L \), averaged over 1000 to 6000 initial configurations.
The insets show the typical 
behavior of the magnetisation \( m_{w} \) at the time of withdrawal of the
field pulse by varying (a) \( h_{p} \) at a fixed \( T \) and \( \Delta t \),
for \( L=50\) and \( 400 \), (b) \( T \) at a fixed
\( h_{p} \) and \( \Delta t \), for \( L=50\) and \(200 \), (c) \( h_{p} \) 
at a fixed \( T \) and \( \Delta t \),
for \( L=50\) and \(200\); \( m_{w}=0 \) at the transition point (from \cite{acbkcpul}).
}{\footnotesize \par}
\vskip 0.1cm

Note that in the low temperature and \( \Delta t \)
regions, there seems to be significant finite size scaling of the
transition (\( m_{w}=0) \) point (see the insets of Fig. 8a, b, c).  In  
Fig. 8d, the finite-size scaling analysis of those data is presented.
For the other cases, there seems to be no significant finite size effect 
on the transition point (cf. insets of Fig. 9a, b, c), indicative of 
a mean-field nature of the transition in this range.
It may be noted that to compare the finite size effects, we normalise the
parameters \(T\) or \(h_p\) by their ranges required for full magnetisation
reversal. In fact, this weak finite size effect for high \(T\) and
\(\Delta t\) regions did not lead to any reasonable value for the fitting 
exponent in the scaling analysis.

For the static transition of the pure two-dimensional Ising system, 
\( U^{*}\simeq 0.6107 \) \cite{binheer,blote}. 
For low temperature (and low \( \Delta t \)) regions of the 
magnetisation-reversal phase boundary, the observed values of \( U^{*} \) (in 
the range 0.42 - 0.46) are considerably lower than the above mentioned value 
for the static transition. There is not enough indication of finite-size 
effect in the \( U^{*} \) value either (cf. \cite{korn1}). This suggests a new
universality class in this range. Also, the finite-size 
scaling study for the effective transition points here (see Fig. 8d) 
gives a correlation length exponent value (\(\nu \simeq  1.4\)) \cite{acbkcpul}
larger than that of the static transition. 
This value of the exponent $\nu$ compares well with that obtained by the 
data collapse ($\nu =2.0\pm 0.5$ in $d=2$) \cite{mish1-1}.
For comparatively higher temperatures (and high \( \Delta t \)), 
the \( U^{*}\simeq 0_{+} \) at the crossing point. Such small value
of the cumulant at the crossing point can hardly be imagined to be a
finite-size effect; it seems unlikely that one would get here also
the same universality class and \( U^{*} \) value will eventually shoot
up to \( U^{*}\simeq 0.44 \) (for larger system sizes), as for the other
range of the transition. On the other hand, such low value of \( U^{*} \)
might indicate a very weak singularity, as indicated by the mean field
calculations \cite{mish1} mentioned in the introduction. In fact,
even for the static transition, as the dimensionality increases, and
the singularity becomes weaker (converging to mean field exponents)
with increasing lattice dimension, the cumulant crossing point \( U^{*} \)
decreases (\( U^{*}\simeq 0.61 \) in \( d=2 \) to \( U^{*}\simeq 0.44 \)
in \( d=4 \)). We believe the mean field transition
behavior here, as mentioned earlier, is even weaker in this dynamic
case as reflected by the value \( U^{*}\simeq 0_{+} \), corresponding
to a logarithmic singularity (as in eqns. (\ref{tau}) and (\ref{xi})).

\section {Summary and conclusions}
In this paper, we review all the aspects studied so far on the dynamic
magnetization-reversal phenomena in the Ising model under a finite-duration
external magnetic field competing with the existing order for $T<T_c^0$.
We have discussed the mean-field equation of motion and its solution, the 
nature of the MF phase boundary and the divergence of the susceptibility and 
relaxation time. The analytic solution of the linearized equation of motion
also showed divergence of susceptibility and relaxation time on the MF phase
boundary \cite{mish1}. The same transition has been studied by Monte Carlo 
simulations in both two and three dimensions. The obtained phase diagram is 
consistent with the classical nucleation theory. The nucleation process is 
initiated by the external magnetic field and depending on the strength of the 
field the system nucleates either through the growth of a single droplet or 
through the growth and subsequent coalescence of many droplets. For 
$h_{p}>h_{DSP}$ the system belongs to the multi-droplet regime and the 
transition is continuous in nature; whereas for \( h_{p}<h_{DSP} \) the system 
goes over to the single-droplet regime where transition is discontinuous. 
Expecting power law behaviour for both \( m_{w} \) and \( \xi  \) in 
multi-droplet regime, the finite size scaling fits give the estimates of the 
critical exponents \( \beta  \) and \( \nu  \) for both \( d=2 \)
and $3$. Unlike in the MF case, where the relaxation time \( \tau_{R}^{MF} \)
shows a logarithmic divergence, $\tau_{R}$ in MC studies falls off exponentially
away from \( m_{w}=0 \) and the divergence in \( \tau _{R} \) comes through
the growth of the prefactor \( \kappa  \) with the system size.

The universality class of the dynamic magnetisation-reversal transition in 2D,
has been studied here using Monte Carlo simulations. Both sequential and
random updating have been used. The phase boundary at any \( T \)
(\(< T_{c}^{0} \)) is obtained first in the \( h_{p}-\Delta t \)
plane. They of course depend on the updating rule. The phase boundaries 
obtained compare well with the nucleation theory estimate 
\(h_p \rm{ln} \Delta t\) = constant along the boundary \cite{pul1,amth}.
Extensive Monte-Carlo studies for the fluctuations in the order parameter 
\(|m_w|\) and internal energies etc. showed prominent divergences along 
the phase boundaries \cite{pul2}. Fourth order cumulant
(\( U \)) of the order parameter distribution is also studied for different
system sizes (upto $L=800$) around the phase boundary region \cite{acbkcpul}.
The crossing point of the cumulant (for different system sizes) gives
the transition point and the value \( U^{*} \) of the cumulant at
the transition point indicates the universality class of the transition.
In the low temperature and low pulse width range, the \( U^{*} \)
value is found to be around 0.44 \cite{acbkcpul} (see Figs. 8a, b, c). The 
prominent discripancy with the \( U^{*} \) value (\(\simeq 0.61\)) for the 
static transition in the same model in two dimensions indicates a new 
universality class for this dynamic transition. Indeed, the finite-size scaling
analysis (Fig. 8d) suggests a different (larger) value of the correlation 
length exponent also. For comparatively higher temperatures and higher pulse 
widths, the $U^{*}$ values are very close to zero (see Fig. 9a, b, c), and the
transitions here seem to fall in a mean-field-like weak-singularity 
universality class similar to that obtained earlier \cite{mish1}, and 
indicated by eqns. (\ref{tau}) and (\ref{xi}).  Here, the finite size effects 
in the order parameter and the transition point are also observed to be 
comparatively weaker (see insets of Fig. 9).

\begin{acknowledgments}
The authors are grateful to Arkajyoti Misra for some useful discussions.
\end{acknowledgments}

\end{document}